\documentclass[prd,preprint,tightenlines,floatfix,preprintnumbers,nofootinbib,eqsecnum,draft]{revtex4}
 \usepackage{color}
      
   \newcommand{\BluTn}[1]{\textcolor{blue}{#1}}     
    \newcommand{\RedTn}[1]{\textcolor{red}{#1}}      
 \usepackage[dvips,final]{graphicx}
  \usepackage{amssymb}
   \usepackage{amsmath}
    \usepackage{epsfig}
     \usepackage{bm}
      \usepackage{pifont}

\newcommand\BraSquare[4]{
 \begin{picture}(0,0)
  \setlength{\unitlength}{1pt}
  \put(#1,#2){\rotatebox{#3}
              {\raisebox{0mm}[0mm][0mm]{
                \makebox[0mm]{$\left.\rule{0mm}{#4pt}\right]$}}}} 
 \end{picture}}                                                   

\begin{document}
\thispagestyle{empty}
\title{Self-consistent Gaussian model of nonperturbative QCD vacuum\\ }
\author{A.~P.~Bakulev}
 \email{bakulev@theor.jinr.ru}
  \affiliation{Bogolyubov Lab. Theor. Phys., JINR, Dubna, 141980 Russia\\ }

\author{A.~V.~Pimikov}
 \email{pimikov@theor.jinr.ru}
  \affiliation{Bogolyubov Lab. Theor. Phys., JINR, Dubna, 141980 Russia\\ }

\vspace {10mm}
\begin{abstract}
\vspace*{+2mm}

We show that the minimal Gaussian model 
 of nonlocal vacuum quark and quark-gluon condensates in QCD 
  generates the non-transversity of vector current correlators. 
We suggest the improved Gaussian model of the nonperturbative QCD vacuum, 
 which respects QCD equations of motion 
  and minimizes the revealed gauge-invariance breakdown. 
We obtain the refined values of pion distribution amplitude (DA) 
 conformal moments 
  $\langle \xi^{2N}\rangle_{\pi}$ ($N=1,..,5$)
   using the improved QCD vacuum model,
    including the inverse moment
     $\langle x^{-1}\rangle_{\pi}$,
      being inaccessible if one uses the standard QCD sum rules.
We construct the allowed region for Gegenbauer coefficients 
 $a_2$ and $a_4$ of the pion DA
   for two values of the QCD vacuum nonlocality parameter,
    $\lambda_q^2=0.4$ and $0.5$~GeV$^2$.
\end{abstract}
\maketitle

\cleardoublepage 
\section{Introduction}
 \label{sec:Intro}
In order to analyze meson distribution amplitudes (DAs) and form factors
the generalization 
of the standard QCD Sum Rules (SRs) approach~\cite{SVZ} 
has been suggested in~\cite{MR86,MR89,BR91,MS93}.
This generalization is based 
on the notion of the nonlocal vacuum condensates (NLC)~\cite{BP81,Gro82,Shu82}
of quark and gluon fields in the nonperturbative QCD vacuum.
The effects of QCD vacuum nonlocality appears 
to be very important
in the pion DA analysis~\cite{Rad94,BM95,BM98,BMS01}.

In this approach we introduce the following gauge-invariant 
quark-antiquark NLCs~\footnote{%
We use the Euclidean interval $x^2 = x_{E}^2 = -x_0^2-\vec{x}^2<0$
and the subscript $E$ will be omitted below for simplicity.}
\begin{eqnarray}
 M_S(x)
  &\equiv&
   \langle{\bar{\psi}(0){\cal E}(0,x)\psi(x)}\rangle
  \ =\ 
   \langle{\bar{\psi}\psi}\rangle
    \int\limits_0^\infty\!\! f_S(\alpha)\,e^{\alpha x^2/4}\,d\alpha\,;
  \label{eq:NLC.S}\\
 M_\mu(x)
  &\equiv&
    \langle{\bar{\psi}(0)\gamma_\mu{\cal E}(0,x)\psi(x)}\rangle
  \ =\
    -ix_\mu A_0
     \int\limits_0^\infty\!\! f_V(\alpha)\,e^{\alpha x^2/4}\,d\alpha\,;
  \label{eq:NLC.V}\\
 {\cal E}(0,x)
  &=& {\cal P}\exp\left[i g \int_0^x A_\mu(\tau) d\tau^\mu\right]\,,
 \label{eq:P.Exp} 
\end{eqnarray}
which are parameterized in the general case by 
distribution functions in virtualities 
$f_S(\alpha)$ and $f_V(\alpha)$,
with $A_0=2\alpha_s\pi\langle{\bar{\psi}\psi}\rangle^2/81$.
Explicit forms of these functions should be taken,
generally speaking, 
from some concrete model of the nonperturbative QCD vacuum.
This can be the exact solution of QCD, 
or some approximation, obtained, for example, in lattice QCD simulation.
In the absence of such a model 
we use the first non-trivial approximation,
taking into account only the finite value of quark momentum distribution 
in the QCD vacuum:
\begin{eqnarray}
 f_S(\alpha) 
  = 
  \delta\left(\alpha-\frac{\lambda_q^2}{2}\right)\,;
  \qquad
 f_V(\alpha)
  =
  {\delta\,}'\!\left(\alpha-\frac{\lambda_\text{V}^2}{2}\right)\,.
  \label{eq:Delta.Ansatz.SV}
\end{eqnarray}
\begin{figure}[b]
 \centerline{\includegraphics[width=0.6\textwidth]{
  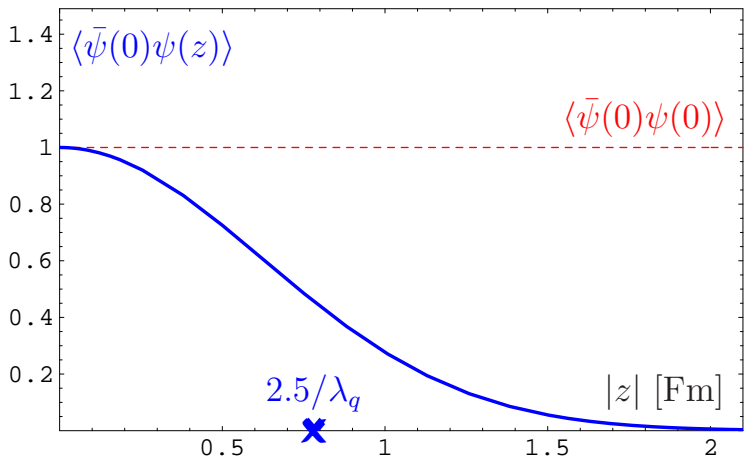}}
   \caption{Quark condensate nonlocality from lattice QCD data 
    of Pisa group~\cite{DDM99,BM02} (solid line).
    Dashed line displays local limit, when quark condensate 
    is constant and does not depend on distance $z$ between quarks.
 \label{fig:NonLocal.Quark.Cond}}
\end{figure}
\\
In this model, 
so-called ``delta-Ansatz'', 
the average virtuality of quarks in the vacuum 
is the single parameter:
\begin{eqnarray} \nonumber 
  \lambda_q^2\equiv 
   \frac{\langle{\bar{\psi}D^2 \psi}\rangle}{\langle{\bar{\psi}\psi}\rangle}\,.
 \end{eqnarray}
We have the following normalization conditions:
\begin{eqnarray} \label{eq:norm}
 \int_{0}^{\infty}\!\!f_{S}(\alpha)\,d\alpha
  = 1\,;\qquad
  \int_{0}^{\infty}\!\!\alpha\,f_{S}(\alpha)\,d\alpha
  = \frac{\lambda_q^2}2\,.
\end{eqnarray}
Higher moments of distribution $f_S(\alpha)$ 
are related with higher dimensional vacuum expectation values (VEVs) 
of quark fields.
Delta-Ansatz (\ref{eq:Delta.Ansatz.SV}) generates Gaussian form 
of NLC in coordinate representation,
\begin{eqnarray}
 M_S(x)
  &=& 
   \langle{\bar{\psi}\psi}\rangle\,
    e^{\lambda_q^2 x^2/8}\,;\qquad
  M_\mu(x)
  \ =\ \frac{i}{4}\,
       x_\mu x^2 A_0\,
    e^{\lambda_\text{V}^2 x^2/8}\,.
 \label{eq:va.SV.Gauss}
\end{eqnarray} 
For this reason below we name it as Gaussian model of NLC. 
The space width of this distribution is approximately
equal to $2.5/\lambda_q$ 
and 
is in a good agreement 
with lattice data 
(in Fig.\,\ref{fig:NonLocal.Quark.Cond} 
 the abscissa of the symbol \BluTn{\ding{54}}
 corresponds to this value).
This model takes into account one but very important property of 
the nonperturbative QCD vacuum --- 
quarks can flow through the vacuum with nonzero momentum $k$
and the average quark virtuality $\langle{k^2}\rangle = \lambda_q^2/2$,
see (\ref{eq:norm}).
It is worth to note here 
that Gaussian asymptotics at large values of $|x|$ 
differs from the anticipated exponential behavior of NLC,
$\sim\exp\left(-\Lambda|x|\right)$.
However, for the moment QCD SRs,
which use averaged with the help of NLC distributions $f(\alpha)$ 
quantities---DA moments~\cite{MR86,MR89}, form factors~\cite{MR90,BR91}---
this wrong asymptotics of NLCs,
as well as the more detailed information about NLC distributions,
is not so important 
(more detailed discussion of this point see in~\cite{BM02}). 

In first papers on NLC SRs~\cite{MR89,BR91} 
it was assumed
that nonlocality parameters of different condensates 
($\lambda_{q}$, $\lambda_\text{V}$ and $\lambda_{\bar{q}Aq}$)
may differ.
In order to simplify the NLC model and 
to diminish the number of parameters
it was suggested and used in subsequent papers~\cite{BM95,BM98,BMS01}
to imply Gaussian model with the single nonlocality parameter 
--- one and the same for scalar and vector NLCs (see~(\ref{eq:Delta.Ansatz.SV})),
and also for quark-gluon-quark (three-local) NLCs:
$\lambda_\text{V}=\lambda_{\bar{q}Aq}=\lambda_q$.
As we will show in this paper, 
such a simplification generates the breakdown of transversal character 
of vector current correlator $\Pi_{\mu\nu}(q)$
and also of Dirac equation for the vector condensate (\ref{eq:NLC.V}).
By this reason 
the construction of a Gaussian NLC model, 
which is minimally consistent with QCD equations of motion 
and minimizes the revealed breakdown of gauge invariance,
seems to be quite reasonable.
We note here that 
the complete restoration of vector current correlator transversity
appeared to be impossible,
because it demands to go outside the frames of Gaussian approximation.

The paper has the following structure.
In the next section we discuss NLCs in QCD:
bi-local 
($\langle{\bar{\psi}(0)\psi(x)}\rangle$ and
 $\langle{\bar{\psi}(0)\gamma_\mu\psi(x)}\rangle$), 
three-local 
($\langle{\bar{\psi}(0)(\gamma_5)\gamma_\mu\hat{A}_\nu(y)\psi(x)}\rangle$)
and four-quark ones
($\langle{\bar{\psi}(0)\psi(y)\bar{\psi}(z)\psi(x)}\rangle$).
Here we obtain an equation
relating bi-local vector NLC with the sum of tree-locals
and following from the QCD Dirac equation for the quark field operator. 
Operator product expansion for the $VV$-correlator $\Pi_{\mu\nu}$,
taking into account nonlocalities of NLCs,
is constructed in the third section.
In the next section we analyze possible delta-Ansatze
and find the best one, called improved Gaussian model,
which minimizes the non-transversal part of the correlator, $\Pi_\text{L}$.
We show in the fifth section the results of NLC QCD SR analysis
for the pion DA with using the improved Gaussian model.
The last section summarizes our conclusions.

\newpage
\section{Basic vacuum condensates}
 \label{sec:QCD.Condensates}
  We use, as usual in QCD SR approach,
  the fixed-point gauge
\begin{eqnarray}
\nonumber
  x^\mu A^{a}_\mu(x) &=& 0\,.
\end{eqnarray}
In this gauge,
the gluon field operator can be expressed in terms of field-strength operators
as follows~\cite{IS82}
\begin{eqnarray}
\nonumber
  A^{a}_\mu(x) &=& x^\nu\!\! \int_0^1\!\!G^{a}_{\nu\mu}(\tau x)\,\tau\,d\tau\,.
\end{eqnarray}
For this reason all Fock--Schwinger strings
$${\cal E}(0,x)\equiv{\cal P}\exp\left[\int_0^x \hat{A}_\mu(z)\,dz^\mu\right]=1$$
if the integration path is a straight line going from $0$ to $x$.

\subsection{Bilocal quark condensates}
 \label{sec:2.Quark.Condensates}
The vacuum expectation value (VEV) of a bilocal quark field operator 
can be written in the general form
\begin{eqnarray}
 \langle{\bar{\psi}^a_A(0)\psi^b_B(x)}\rangle
      = \frac{\delta^{ab}}{N_c}
      \int\limits_0^\infty  
      \left\{\frac{\delta_{AB}}{4}\,
              \langle{\bar{\psi}\psi}\rangle\,
              f_S(\alpha)
           - \frac{\widehat{x}_{BA}}{4}\, 
              iA_0\,
               f_V(\alpha)
      \right\}
      e^{\alpha x^2/4}\,d\alpha\,,
       \label{eq:va1}
\end{eqnarray}
where $A_0={2\alpha_s\pi\langle{\bar{\psi}\psi}\rangle^2}/{81}$, 
and  functions $f_S(\alpha)$ и $f_V(\alpha)$ 
parameterize 
the scalar and vector condensates, respectively.
The transition to the local case is evident
\begin{eqnarray}\nonumber%
 f^\text{loc}_S(\alpha)=\delta(\alpha)\,;\qquad
 f^\text{loc}_V(\alpha)=\delta\,'(\alpha)\,.
\end{eqnarray}

\subsection{Trilocal quark-gluon condensates}
 \label{sec:3-Local.Condensates}
It is convenient to term the quark-gluon-antiquark  condensate 
in the fixed-point gauge 
by introducing three scalar functions
$\overline{M}_{1,2,3}(x^2,y^2,z^2)$~\cite{MR89,BR91,BM98}:
\begin{eqnarray}
M_{\mu\nu}(x,y)&\equiv&
          \langle{\bar{\psi}(0)\gamma_\mu(-g\widehat{A}_\nu(y))\psi(x)}\rangle
       =\nonumber\\
      &=&
      (y_\mu x_\nu-g_{\mu\nu}(xy))\overline{M}_1(x^2,y^2,(x-y)^2)\nonumber\\
      &+&
      (y_\mu y_\nu-g_{\mu\nu}y^2)\overline{M}_2(x^2,y^2,(x-y)^2)\,;
 \label{eq:M.mu.nu}\\
M_{5\mu\nu}(x,y)&\equiv&
          \langle{\bar{\psi}(0)\gamma_5\gamma_\mu(-g\widehat{A}_\nu(y))\psi(x)}\rangle
      = 
      i\varepsilon_{\mu\nu yx}\overline{M}_3(x^2,y^2,(x-y)^2)\,,\nonumber
\end{eqnarray}
where $A_i=\{-\frac{3}{2},2,\frac{3}{2}\}A_0$ and 
$\overline{M}_{1,2,3}(x^2,y^2,z^2)$ can be parameterized as
\begin{eqnarray}
\overline{M}_i(x^2,y^2,(x-y)^2)=
            A_i\int\limits_{\!0}^{~\infty}\!\!\!\!\int\limits_{\!0}\!\!\!\!\int\limits_{\!0}^{\infty}\!\!
            d\alpha_1\, d\alpha_2\, d\alpha_3 f_i(\alpha_1,\alpha_2,\alpha_3)\,
            e^{\left(\alpha_1x^2+\alpha_2y^2+\alpha_3(x-y)^2\right)/4}\,.
 \nonumber
\end{eqnarray}
We prefer to use here the hypothesis
that quark and anti-quark are interchangeable 
in $\bar{q}Gq$-condensate,
which means that
\begin{eqnarray}
 \label{eq:sym.a2-a3}
  f_i(\alpha_1,\alpha_2,\alpha_3)
  = f_i(\alpha_1,\alpha_3,\alpha_2)\,. 
\end{eqnarray}
Transition to the local case for these functions 
is defined as follows
$$f^\text{loc}_i(\sigma,\rho,\tau)
  = \delta(\sigma)\,\delta(\rho)\,\delta(\tau)\,.$$

\subsection{QCD equation of motion and nonlocal condensates}
 \label{sec:Eq.Motion}
Dirac equation for the quark field operator in massless QCD
$$\hat{\nabla}\psi(x)=0$$
allows us to write down immediately 
the equation of motion 
for the splitted quark current
$j_\mu(x) = \bar{\psi}(0)\gamma_\mu \psi(x)$
\begin{eqnarray}
  \nabla^\mu j_\mu(x)  &=& 0\,,
 \nonumber
\end{eqnarray}
where $\nabla_\mu^{AB}$ is covariant derivative.
If we sandwich this operator equation between physical QCD vacuum states
than we obtain the equation for condensates:
\begin{subequations}
 \label{eq:Eq.Mot.QgAQ}
  \begin{eqnarray}
   \partial^\mu \langle{0|\bar{\psi}(0)\gamma_\mu \psi(x)|0}\rangle
    &=& i\langle{0|\bar{\psi}(0) \gamma_\mu g\hat{A}^\mu(x) \psi(x)|0}
         \rangle\,;\\
   \partial^\mu M_\mu(x)
    &=&-i M_{\mu\phantom{\mu}}^{\phantom{\mu}\mu}(x,x)\,.
  \end{eqnarray}
\end{subequations}
Let us first consider the left-hand side (l.h.s.) part of this relation.
By substituting (\ref{eq:va1}) and  delta-Ansatz (\ref{eq:Delta.Ansatz.SV}) 
with $\lambda_V^2/2=\Lambda$
into this part 
we obtain
\begin{eqnarray}
 \partial^\mu M_\mu(x)  
  = +\frac{iA_0\,x^2}{2}\left[3 + \frac{\Lambda x^2}{4}\right]\,
       e^{\Lambda x^2/4}\,.
 \nonumber
\end{eqnarray}
The right-hand side (r.h.s) part of (\ref{eq:Eq.Mot.QgAQ}) 
can be rewritten by using (\ref{eq:M.mu.nu}):
\begin{eqnarray}
 -i M_{\mu\phantom{\mu}}^{\phantom{\mu}\mu}(x,x)
 = +\frac{iA_0\,x^2}{2}\int\limits_{\!0}^{\,\infty}\!
     \langle{\langle{12 f_2- 9f_1}\rangle}\rangle(\alpha)\,
      e^{\alpha x^2/4}\,d\alpha\,,
 \nonumber
\end{eqnarray}
where we defined the averaging $\langle{\langle{\ldots}\rangle}\rangle$
as
\begin{eqnarray}\nonumber%
 \langle{\langle{f_i}\rangle}\rangle(\alpha)
  \equiv \int\limits_{\!0}^{1}\!\!\alpha\,dx\!\!\int\limits_{\!0}^{\,\infty}\!\!\!\,
          d\alpha_3\,
           f_i\left(x\alpha,(1-x)\alpha,\alpha_3\right)\,.
\end{eqnarray}
Using (\ref{eq:Eq.Mot.QgAQ}), we get
\begin{eqnarray}
 \int\limits_{\!0}^{\,\infty}\!
  \langle{\langle{12 f_2- 9f_1}\rangle}\rangle(\alpha)\,
   e^{\alpha x^2/4}\,d\alpha
 \ =\
  \left[3 + \frac{\Lambda x^2}{4}\right]\,e^{\Lambda x^2/4}\,.
  \label{eq:Eq.Mot.QgAQ.Red}
\end{eqnarray}
We see immediately 
that if one uses the minimal delta-Ansatz 
for $f_1$ and $f_2$ functions
\begin{eqnarray}
 \label{eq:3L.D.A.Naive}
  f_i^\text{min}\left(\alpha_1,\alpha_2,\alpha_3\right)
  &=& \delta\left(\alpha_1-x_{i}\Lambda\right)
       \delta\left(\alpha_2-y_{i}\Lambda\right)
        \delta\left(\alpha_3-z_{i}\Lambda\right)\,,
\end{eqnarray}
then,
in order to have the same exponential in both sides
of Eq.\ (\ref{eq:Eq.Mot.QgAQ.Red}),
we have to set
\begin{eqnarray}
 \label{eq:x_i+y_i}
  x_{i}+y_{i} &=& 1\,.
\end{eqnarray}
But this condition is not sufficient 
to fulfill Eq.\ (\ref{eq:Eq.Mot.QgAQ.Red}):
the minimal Ansatz generates only the first term 
in the square brackets in the r.h.s of (\ref{eq:Eq.Mot.QgAQ.Red}),
namely $3\cdot\exp\left(\Lambda x^2/4\right)$.
To cure this deficiency,
we suggest here to use the improved delta-Ansatz:
\begin{eqnarray}
 \label{eq:3L.D.A.Deriv}
  f_i\left(\alpha_1,\alpha_2,\alpha_3\right)
  &=& \left(1 + X_{i}\partial_{x_i}
              + Y_{i}\partial_{y_i}
              + Z_{i}\partial_{z_i}\right)
       \delta\left(\alpha_1-x_{i}\Lambda\right)\,
        \delta\left(\alpha_2-y_{i}\Lambda\right)\,
         \delta\left(\alpha_3-z_{i}\Lambda\right)\,.~~~
\end{eqnarray}
Then, in addition to condition (\ref{eq:x_i+y_i}),
we obtain the condition for coefficients $X_{i}$ and $Y_{i}$:
\begin{eqnarray}
 \label{eq:3L.D.A.Rule4}
  12\,\left(X_{2} + Y_{2}\right)
  - 9\,\left(X_{1} + Y_{1}\right)
  = 1 \,.~
\end{eqnarray}

\subsection{Four-quark condensates}
 \label{sec:4-Quark.Condensates}
Vacuum condensates of 4-quarks operators are usually transformed 
to the product of two scalar quark condensates 
by means of the Hypothesis of Vacuum Dominance (HVD)\footnote{%
For shortness we consider operators $A$ and $B$,
which include also color matrixes $t^a$ and $t^b$.}
\begin{eqnarray}
 \langle{\bar{\psi}(0) A \psi(y)
          \bar{\psi}(z) B \psi(x)}\rangle
 \cong  \left(\frac{-\text{Tr\,} AB}{16\,N_c^2}\right) 
          M_\text{S}\left(x^2\right)
           M_\text{S}\left((z-y)^2\right)\,,
   \label{eq:GVD.NL.Nai}
\end{eqnarray}
Due to decay of correlations at large distances,
say,
when $y^2$ and $(z-x)^2$ are much larger 
than the characteristic scale 
of QCD vacuum nonlocality, $1/\lambda_q^2\sim(0.3~\text{Fm})^2$,
HVD should work well.
In the opposite case,
namely, when $(z-x)^2\ll1/\lambda_q^2$ or $y^2\ll1/\lambda_q^2$,
we should have the HVD breakdown,
which is related with true 4-quarks correlations.
To take this breakdown into consideration
we can add the form factor $\Phi_{4}\left(y^2+(x-z)^2\right)$,
accounting for the separation of quark pairs $(0,x)$ and $(z,y)$:
\begin{eqnarray}
 \langle{\bar{\psi}(0) A \psi(y)
          \bar{\psi}(z) B \psi(x)}\rangle
 \cong  \left(\frac{-\text{Tr\,} AB}{16\,N_c^2}\right) 
         M_\text{S}\left(x^2\right)
          M_\text{S}\left((z-y)^2\right)
          \left[1+\Phi_{4}\left(y^2+(x-z)^2\right)\right],~
   \nonumber
\end{eqnarray}
where $\Phi_{4}(x^2)$ decreases fast for $x^2\gg1/\lambda_q^2$.
This modification can be done, but it does not appear to be very important.
We will consider the influence of this modification in a separate paper.
We suppose here, that $\Phi_{4}(x^2)=0$.

 \section{Operator product expansion of vector current correlator}
  \label{sec:VV.Correlator}
Consider now the correlator 
\begin{eqnarray}
 \Pi^N_{\mu\nu}
    =i\,\int\!\!d^4x\, e^{iqx} 
      \langle{0\big|T\left[J^{N}_\mu(0)J_\nu^{+}(x)\right]\big|0}\rangle\,,
       \label{eq:Correl.mu.nu.N}
\end{eqnarray}
of two vector currents corresponding to charged $\rho$ meson
\begin{eqnarray}
 \nonumber
  J^N_\mu(0) 
  = \bar{d}(0)\gamma_\mu \left(-in\nabla_0\right)^N u(0)\,;
  \qquad
  J_\nu^{+}(x) = \bar{u}(x)\gamma_\nu d(x)\,.
\end{eqnarray}
In the first current we have the composite operator $\left(-in\nabla_0\right)^N$.
Its action on the quark field is defined as
\begin{eqnarray}
 \nonumber
  \left(-in\nabla_0\right)^N u(0)
   \equiv 
    \left.\bar{d}(0)\gamma_\mu \left(-in\nabla_y\right)^N u(y)\right|_{y=0}\,,
\end{eqnarray}
where $n$ is an arbitrary light-like vector, $n^2=0$,
such that $nq\neq0$.

For shortness, we will write below $\Pi_{\mu\nu}^N$ in place of $\Pi_{\mu\nu}^N(q)$.
Note that the correlator $\Pi_{\mu\nu}^N$ depends on two vectors $q$ and $n$.
This dependency allows to write correlator in term s of the following Lorentz structures
\begin{eqnarray}
 \label{eq:Pi.Lorenz.A.E}
  \Pi_{\mu\nu}^N
   &=& A_N\,q_\mu\,q_\nu
     + B_N\,g_{\mu\nu}\,q^2
     + C_N\,\frac{n_\mu\,n_\nu}{nq^2}\,q^4
     + D_N\,\frac{q_\mu\,n_\nu}{nq}\,q^2
     + E_N\,\frac{n_\mu\,q_\nu}{nq}\,q^2
\end{eqnarray}
and
\begin{eqnarray}
 \label{eq:Pi.Lorenz.T.L}
  \Pi_{\mu\nu}^N
   &=& \Pi_{\text{T}_1}^N\,
        \left[q_\mu\,q_\nu-g_{\mu\nu}\,q^2\right]
     + \Pi_{\text{T}_2}^N\,
        \left[g_{\mu\nu}\,q^2 
           + \left(\frac{n_\mu\,n_\nu}{nq^2}\,q^2
                 - \frac{q_\mu\,n_\nu+n_\mu\,q_\nu}{nq}
             \right)\,q^2
        \right]
 \nonumber\\
   &+& \Pi_{\text{T}_3}^N\,
        \left[q_\mu\,q_\nu - \frac{q_\mu\,n_\nu}{nq}\,q^2\right]
     + \Pi_\text{L}^N\,
        \left[\frac{q_\mu\,n_\nu+n_\mu\,q_\nu}{nq}\,q^2\right]
     + \Pi_\text{LL}^N\,
        \frac{n_\mu\,n_\nu}{nq^2}\,q^4\,.
\end{eqnarray}
Lorentz-invariant structures  $A_N, \ldots, E_N$
and $\Pi_\text{T$_i$}^N$, $\Pi_\text{L}^N$, $\Pi_\text{LL}^N$
are connected by the simple algebraic relations
\begin{subequations}
\label{eq:T.L_A.E}
\begin{eqnarray}
  \Pi_{\text{T}_1}^N
   = A_N + D_N - E_N\,;
  \quad
  \Pi_{\text{T}_2}^N
   = A_N + B_N + D_N - E_N\,;
  \quad
  \Pi_{\text{T}_3}^N
   = E_N - D_N\,;\\
  \Pi_\text{L}^N
   = A_N + B_N + D_N\,;  
  \qquad
  \Pi_\text{LL}^N
   = C_N + E_N - A_N - B_N - D_N\,.
 ~~~~~~~~~~~
\end{eqnarray}
\end{subequations}
Taking into account conservation of vector current $J_{\nu}(x)$
we get:
\begin{eqnarray}
 \label{eq:Pi.q.nu}
  q^\nu\,\Pi_{\mu\nu}^N 
  = q^\mu\,q^2\,\Pi_\text{L}^N 
  + \frac{n^\mu\,q^4}{nq}\,
     \left(\Pi_\text{L}^N + \Pi_\text{LL}^N \right)
  = 0
\end{eqnarray}
or in terms of $A_N, \ldots, E_N$,
\begin{eqnarray}
 \label{eq:AE.q.nu}
  q^\nu\,\Pi_{\mu\nu}^N 
  = q^\mu\,q^2\,\left(A_N + B_N + D_N\right)
  + \frac{n^\mu\,q^4}{nq}\,
     \left(C_N + E_N\right)
  = 0\,.
\end{eqnarray}
\begin{figure}[b]
 \centerline{\includegraphics[width=0.4\textwidth]{
             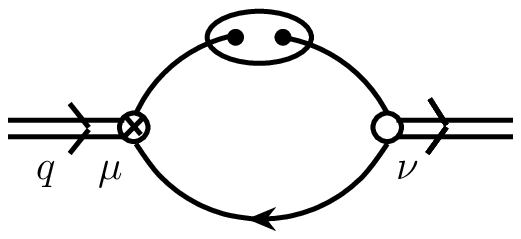}~~~%
             \begin{minipage}{0.4\textwidth}\vspace*{-69pt}
             \centerline{\includegraphics[width=\textwidth]{
             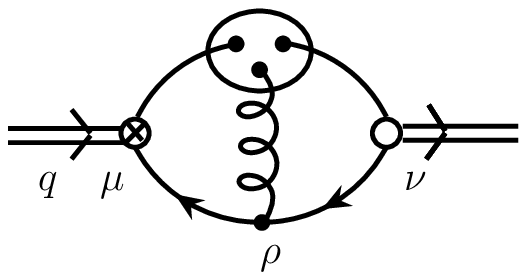}}
             \end{minipage}}
  \caption{\label{fig:NLC1} 
   Vector quark-quark ($\Delta_\text{2V}\Pi^N_{\mu\nu}$, left) 
   and quark-gluon-antiquark ($\Delta_{\bar{q}Aq}\Pi^N_{\mu\nu}$, right)
   condensates contributions to the correlator $\Pi^N_{\mu\nu}$.}
  \end{figure}
We will analyze the $\Pi_\text{L}^N$ structure,
which can be obtained using the projector $n^\mu q^\nu/(nq)$.
This structure is the most important one,
because it distorts just the coefficient $A_N$.
And namely this coefficient is responsible 
for distribution amplitudes (DAs) of the leading twist.

\begin{figure}[t]
 \centerline{\includegraphics[width=0.4\textwidth]{
             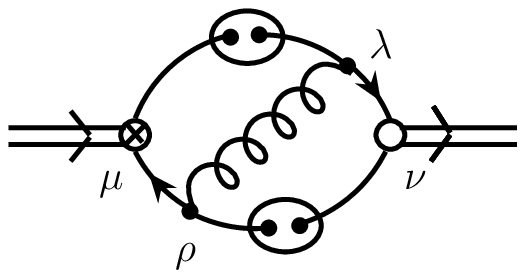}~~~%
             \begin{minipage}{0.4\textwidth}\vspace*{-99pt}
             \centerline{\includegraphics[width=\textwidth]{
             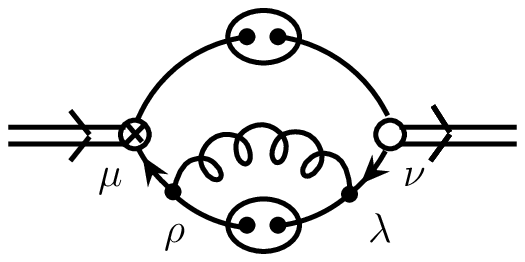}}
             \end{minipage}}
  \caption{\label{fig:NLC2} 
   Four-quark condensates contributions to the correlator $\Pi^N_{\mu\nu}$:
   $\Delta_\text{4q1}\Pi^N_{\mu\nu}$ (left) and 
   $\Delta_\text{4q2}\Pi^N_{\mu\nu}$ (right).}
\end{figure}

All  $O(\alpha_s\langle{\bar{\psi}\psi}\rangle^2)$-terms in $\Pi_{\mu\nu}^N$ 
are generated 
by bilocal vector, 
quark-gluon-antiquark (see Fig.\,\ref{fig:NLC1}),
and 4-quarks condensates (see Fig.\ \ref{fig:NLC2}):
\begin{eqnarray}
 \label{eq:Pi.mu.nu.N}
  \Pi_{\mu\nu}^N  =  \Delta_\text{2V}\Pi^N_{\mu\nu}
                  + \Delta_{\bar{q}Aq}\Pi^N_{\mu\nu}
                  + \Delta_\text{4Q$_1$}\Pi^N_{\mu\nu}
                  + \Delta_\text{4Q$_2$}\Pi^N_{\mu\nu}
                  + (\text{M.~C.})\,.
\end{eqnarray}
M.~C. means terms due to mirror-conjugated diagrams:
for example, in Fig.\,\ref{fig:NLC1} they correspond to diagrams,
in which NLC are inserted in the bottom line 
instead of the top one.

We are interested in the quantities 
corresponding to the non-transversal structure $\Pi_\text{L}^N$:
$$\Delta_{k}\Pi_\text{L}^N(M^2)
   \equiv \frac{M^4}{2A_0}\,
    \widehat{B}_{-q^2\rightarrow M^2}\,
     \frac{\Delta_{k}\Pi^N_{\mu\nu}\,n^\mu q^\nu}{nq}
   = \int\limits_0^1\!x^N\varphi_{k,\,L}(x,M^2)\,dx$$ 
with $k=2$V, $\bar{q}Aq$, 4Q$_1$ и 4Q$_2$.
Here, in close analogy with QCD SR approach,
we work with Borelized quantities,
which are obtained after Borel transformation
$\widehat{B}_{-q^2\rightarrow M^2}$.
Using our parameterizations of vacuum condensates (\ref{eq:va1}), (\ref{eq:M.mu.nu}),
we obtain for these terms:
\begin{subequations}
\begin{eqnarray}
 \label{eq:2.V.L}
  \varphi_{2V,\,L}(x,M^2)
       &=& - M^4\,x\,\left(f_V\left(M^2 \bar{x}\right)
                      -M^2\,\bar{x}\,f_V'\left(M^2 \bar{x}\right)
                     \right)
   \,;~~~\\
 \label{eq:q.A.q.L}
  \Delta_{\bar{q}Aq}\Pi_\text{L}^N(M^2)
       &=& \sum\limits_{i=1}^{3}   
            \frac{2A_i}{A_0}
             \mathop{\int\!\!\!\!\int\!\!\!\!\int}^{~~\infty}_{0\,0\,0~~}\!\!
            d\alpha_1\, d\alpha_2\, d\alpha_3\,
            f_i(\alpha_1,\alpha_2,\alpha_3)\,
            \frac{G_i(\bar{\Delta}_1-\Delta _2)^N+H_i \bar{\Delta}_1^{N+2}}
                 {(N+2)\,\bar{\Delta}_1^3 \Delta _2^3}
   \,;~~~~~\\
 \label{eq:4.q.1.L}
  \Delta_\text{4Q$_1$}\Pi_\text{L}^N(M^2)
       &=& 18\,
           \frac{\left(\log\left(\bar{\Delta}\right)F_1+F_2\right)\bar{\Delta}^{N+2}+F_3}
                {(N+2)^2(N+3)\Delta\bar{\Delta}^2}
   \,;\\
 \label{eq:4.q.2.L}
  \varphi_\text{4Q$_2$,\,L}(x,M^2)
       &=& 36\,M^2\,\frac{f_S\left(M^2 \bar{x}\right)}{x}
   \,,~~~
\end{eqnarray}
\end{subequations}
where $\Delta=\Lambda_S/M^2$, $\bar{\Delta}=1-\Delta$,  
    $\Delta_i=\alpha_i/M^2$, $\bar{\Delta}_1=1-\Delta_1$.
Explicit form of functions $F_i$, $G_i$, and $H_i$ 
are given 
in Appendix~\ref{App:A.1}. 
The term $\Delta_\text{4Q$_1$}\Pi_\text{L}^N(M^2)$
is written for Ansatz (\ref{eq:Delta.Ansatz.SV}).

As we stated above, 
the most important for DAs of the leading twist
is the coefficient $A_N$ in front of the structure $q_\mu\,q_\nu$
in (\ref{eq:Pi.Lorenz.A.E}).
Having in mind further applications for meson DAs,
we calculate the corresponding contributions 
($k=2$V, $\bar{q}Aq$, 4Q$_1$ and 4Q$_2$):
\begin{eqnarray}
 \label{eq:Delta.Pi.T.k}
  \Delta_{k}\Pi_\text{T}^N(M^2)
   \equiv \frac{M^6}{2A_0}\,
    \widehat{B}_{-q^2\rightarrow M^2}
     \frac{\Delta_{k}\Pi^N_{\mu\nu}\,n^\mu n^\nu}{nq^2}
   = \int\limits_0^1\!x^N\varphi_{k,\,T}(x,\,M^2)\,dx\,.
\end{eqnarray}
We obtain the following expressions
\begin{eqnarray}
 \varphi_{2\text{V},\,T}(x,\,M^2)
   &=& 2\,M^4\,x\,f_V(M^2\bar{x})
  \,;~~~\\
 \varphi_{\bar{q}Aq,\,T}(x,\,M^2)
   &=& \frac{4 A_i}{A_0}
        \mathop{\int\!\!\!\!\int\!\!\!\!\int}^{~~\infty}_{0\,0\,0~~}\!\!
         d\alpha_1\, d\alpha_2\, d\alpha_3\,
          f_i(\alpha_1,\alpha_2,\alpha_3)\,
           \widetilde{\varphi}_{i}(\alpha_1,\alpha_2,\alpha_3,M^2)
  \,;~~~\\
 \varphi_{\text{4Q$_1$},\,T}(x,\,M^2)
   &=& 36\mathop{\int\!\!\!\!\int}^{~~\infty}_{0\,0~}\!\!
         d\alpha_1\, d\alpha_2\,
          f_S(\alpha_1)\,
          f_S(\alpha_2)\,
           \widetilde{\varphi}(\alpha_1,\alpha_2,M^2)
  \,;~~~\\
 \varphi_{\text{4Q$_2$},\,T}(x,\,M^2) 
   &=& 0\,,
\end{eqnarray}
where functions
$\widetilde{\varphi}_i(\alpha_1,\alpha_2,\alpha_3,M^2)$
and 
$\widetilde{\varphi}(\alpha_1,\alpha_2,M^2)$ are given 
in the explicit form 
in Appendix~\ref{App:A.1}.

\section{Analysis of Gaussian models}
 \label{sec:Analysis.Delta-Ansatze}
Vector current conservation in QCD claims 
for the transversity (with respect to the index $\nu$)
of the sum of contributions all quark condensates:
\begin{eqnarray}
 \label{eq:2vplus3L.equiv.0}
   \Delta\Pi_\text{L}^N
    \equiv
     \Delta_\text{2V}\Pi^N_\text{L}
   + \Delta_{\bar{q}Aq}\Pi^N_\text{L}
   + \Delta_\text{4Q$_1$}\Pi^N_\text{L}
   + \Delta_\text{4Q$_2$}\Pi^N_\text{L}
   + (\text{M.~C.})
   = 0\,.~
\end{eqnarray}
Note here that since we study Gaussian model
based on delta-Ansatz (\ref{eq:Delta.Ansatz.SV}), 
(\ref{eq:3L.D.A.Deriv}),
the sum $\Delta\Pi_\text{L}^N$ can not be equal zero
exactly.
The reason is very simple --- we insert the Gaussian behavior by hands.
So, the only thing we can hope to realize, 
is to minimize $\big|\Delta\Pi_\text{L}^N\big|$ 
by the special choice of the Ansatz's parameters.
More precisely, 
we are interested in minimization 
of conformal moments $\Delta\langle\xi^{2N}\rangle_\text{L}$,
which are used in the QCD SR analysis of meson DAs.
Relations between moments 
$\Delta\langle\xi^{2N}\rangle_\text{L}$ and 
$\Delta\Pi_\text{L}^N$ 
are considered in Appendix \ref{App:A.2}.

In order to find these values of our parameters $\left\{X_i\right\}$
we introduce the following optimization function 
($\Delta\equiv\lambda_{q}^2/(2M^2)$)
\begin{eqnarray}
 \nonumber
  \Phi_K\left(\left\{X_i\right\}\right) 
   &=& \sum_{N=0}^{K}w_N\,
        \langle\langle
          \left|\Delta\langle\xi^{2N}\rangle_\text{L}
           \left(\Delta;\left\{X_i\right\}\right)
          \right|^2
        \rangle\rangle\,;\\
\nonumber
  \langle\langle F\left(\Delta\right)\rangle\rangle
   &\equiv&
     \frac{1}{17}\,\sum_{j=1}^{17}
      F\left(\Delta=0.024\cdot j\right)\,,
\end{eqnarray}
summing up ``norms'' of first nontrivial $K$ moments
$\Delta\langle\xi^{n}\rangle_\text{L}\left(\Delta;\left\{X_i\right\}\right)$
with $n=0$, $2$, $\ldots\ $, $2K$.
To define the corresponding ``norm'' of function $F(\Delta)$,
$\langle\langle F\left(\Delta\right)\rangle\rangle$,
we integrate numerically in $\Delta\in[0-0.45]$,
because just this interval of $\Delta$ values is physically important 
in QCD SRs.
The weights $w_N$ are specified by the corresponding norms
in the minimal Ansatz case~\cite{BM98,BMS01}:
\begin{eqnarray}
 \nonumber
  \Phi_{2N}^\text{min}
   &=& \langle\langle
          \left|\Delta\langle\xi^{2N}\rangle_\text{L}
           \left(\Delta;\left\{X_v=1, X_i=Y_i=Z_1=0, x_i=y_i=z_i=1\right\}\right)
          \right|^2
        \rangle\rangle\,.
\end{eqnarray}
We introduce these weights in order 
to normalize contributions of different moments to the whole sum 
and 
to make these contributions of the same order of magnitude.
For this reason we define $w_N$ as follows:
\begin{eqnarray}
 \nonumber
  w_N
   &=& \frac{\Phi_{0}^\text{min}}{\Phi_{2N}^\text{min}}\,.
\end{eqnarray}
This receipt gives us in the minimal Ansatz case
equal to unity contributions of all moments to the optimization function
$\Phi_K\left(\left\{X_v=1, X_i=Y_i=Z_1=0, x_i=y_i=z_i=1\right\}\right)=K+1$.
Numerically, we use $K=\,5$ and find
\begin{eqnarray}
 \nonumber
 w_0=1\,;\quad 
 w_2=13\,;\quad 
 w_4=29\,;\quad 
 w_6=45\,;\quad 
 w_8=55\,;\quad 
 w_{10}=59\,.
\end{eqnarray}
\begin{figure}[t]
 \centerline{\includegraphics[width=0.32\textwidth]{
             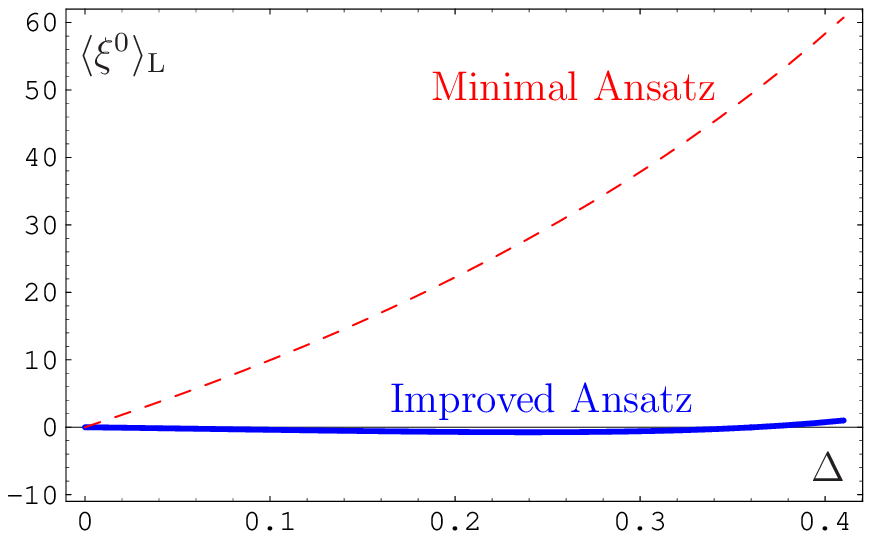}~%
             \includegraphics[width=0.32\textwidth]{
             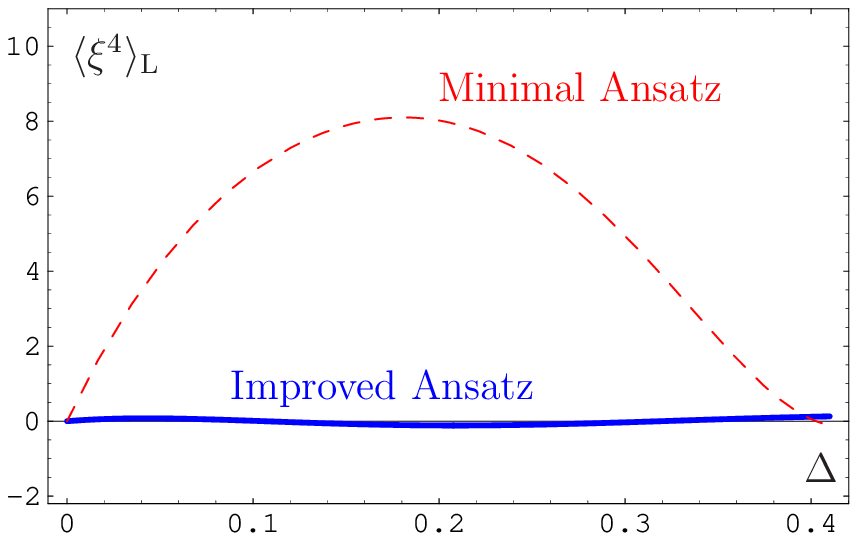}~%
             \includegraphics[width=0.32\textwidth]{
             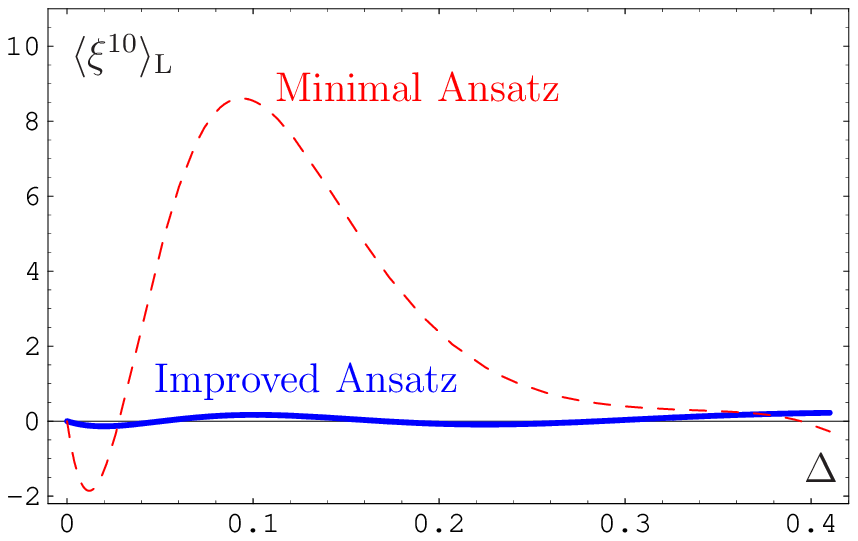}}
  \caption{\label{fig:gra.Imp.Naiv.00} 
            We show functions $\Delta\langle\xi^{2N}\rangle_\text{L}(\Delta)$ 
            with $N=0, 4, 10$  for the improved NLC model (\ref{eq:Ansatz.Xv1.W})
            (solid line)  in comparison with ones, corresponding 
            to the minimal NLC model  (dashed line).}
\end{figure}
\begin{figure}[b]
 \centerline{\includegraphics[width=0.45\textwidth]{
             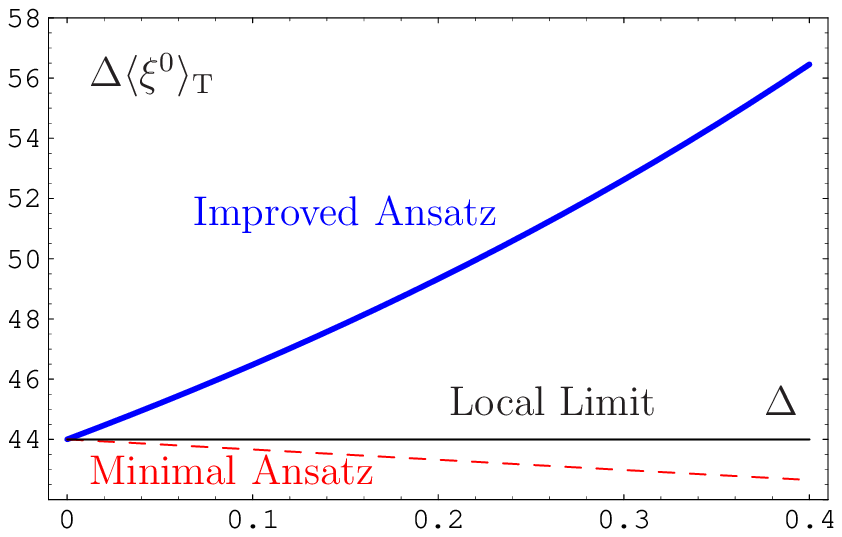}~~~%
             \includegraphics[width=0.45\textwidth]{
             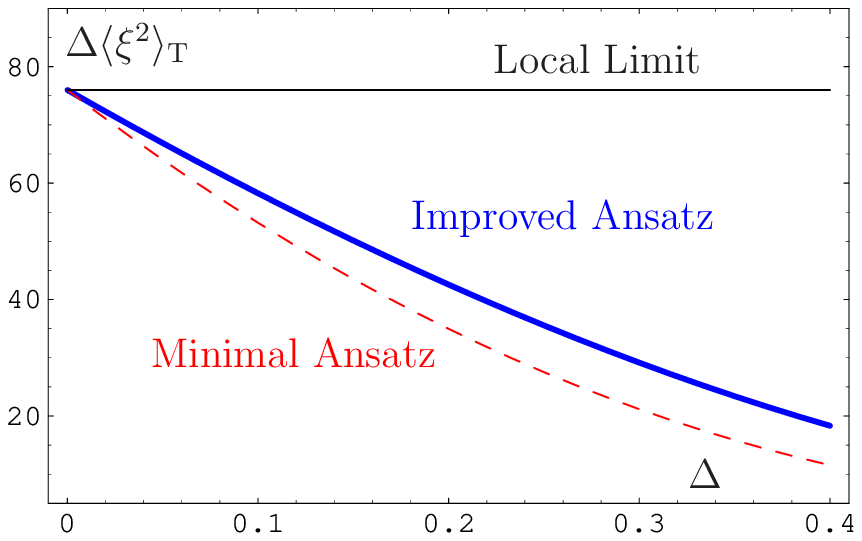}}
  \caption{\label{fig:gra.Imp.Naiv.00.T} 
  We show functions $\Delta\langle\xi^{0}\rangle_\text{T}$ (left panel)
  and $\Delta\langle\xi^{2}\rangle_\text{T}$ (right panel)
  for the improved NLC model (\ref{eq:Ansatz.Xv1.W}) (solid line) 
  in comparison with ones, corresponding to the minimal NLC model
  (dashed line).}
\end{figure}

Consider now the set of available parameters $\left\{X_i\right\}$
in our improved model.
We apply delta Ansatz (\ref{eq:Delta.Ansatz.SV})
with one parameter $X_v$,
relating nonlocalities in vector and scalar quark condensates:
\begin{eqnarray}
 \label{eq:X.V}
  \lambda_\text{V}^2
   = X_v\,\lambda_q^2\,.
\end{eqnarray}
For the quark-gluon-quark condensate we use Ansatz (\ref{eq:3L.D.A.Deriv}) 
with $\Lambda=X_v\lambda_q^2/2$ 
and apply condition (\ref{eq:sym.a2-a3}):
\begin{eqnarray}
 \nonumber
  \quad Z_{i}=Y_{i}\,;
  \quad x_{i}=x\,;
  \quad y_{i}=z_{i}=1-x\,;\quad
  (i=1,\ 2,\ 3)\,.
\end{eqnarray}
These parameters are not independent
due to Eq.\ (\ref{eq:3L.D.A.Rule4}),
derived from the QCD equation of motion (\ref{eq:Eq.Mot.QgAQ}).
After taking into account all mentioned relations
we have the following 7 parameters: 
$x$, $X_1$, $X_2$, $X_3$, $Y_1$, $Y_3$ and $X_v$.
Minimization of the function $\Phi_5\left(x, X_1, X_2, X_3, Y_1, Y_3, X_v\right)$
gives us the following set of parameters
\begin{eqnarray}\nonumber
 X_1 &=& +0.082\,;\quad Y_1 = Z_1 = -2.243\,;\quad x_1=x_2=x_3=x= 0.788\,;\quad X_v = 1.00\,;~~~\\
 X_2 &=& -1.298\,;\quad Y_2 = Z_2 = -0.239\,;\quad y_1=y_2=y_3=1-x= 0.212\,;\label{eq:Ansatz.Xv1.W}\\
 X_3 &=& +1.775\,;\quad Y_3 = Z_3 = -3.166\,;\quad z_1=z_2=z_3=1-x= 0.212\,.\nonumber              
\end{eqnarray}
We will consider this set as the basic parameter set of the improved Gaussian model.

To illustrate the quality of the improved Ansatz
we show in Fig.\ \ref{fig:gra.Imp.Naiv.00}
plots of the functions $\Delta\langle\xi^{2N}\rangle_\text{L}(\Delta)$ 
with $N=0, 2, 5$ (solid lines)
in comparison with corresponding quantities for the minimal Ansatz (dashed lines).
As is clearly seen from this comparisons, 
the improved Ansatz (\ref{eq:Ansatz.Xv1.W}) strongly suppresses 
the absolute values of non-transverse conformal moments 
$\Delta\langle\xi^{2N}\rangle_\text{L}$,
i.~e. takes the vector correlator transversity into account much better.

In Fig. \ref{fig:gra.Imp.Naiv.00.T} we also show 
the moments $\Delta\langle\xi^{2N}\rangle_\text{T}$ 
with $N=0$ and $N=1$
for the improved Gaussian model 
in comparison with results for the minimal model~\cite{BMS01}.

\section{Pion distribution amplitude}
 \label{sec:Pion.DA}
\begin{figure}[h]
 \centerline{\includegraphics[width=0.47\textwidth]{
              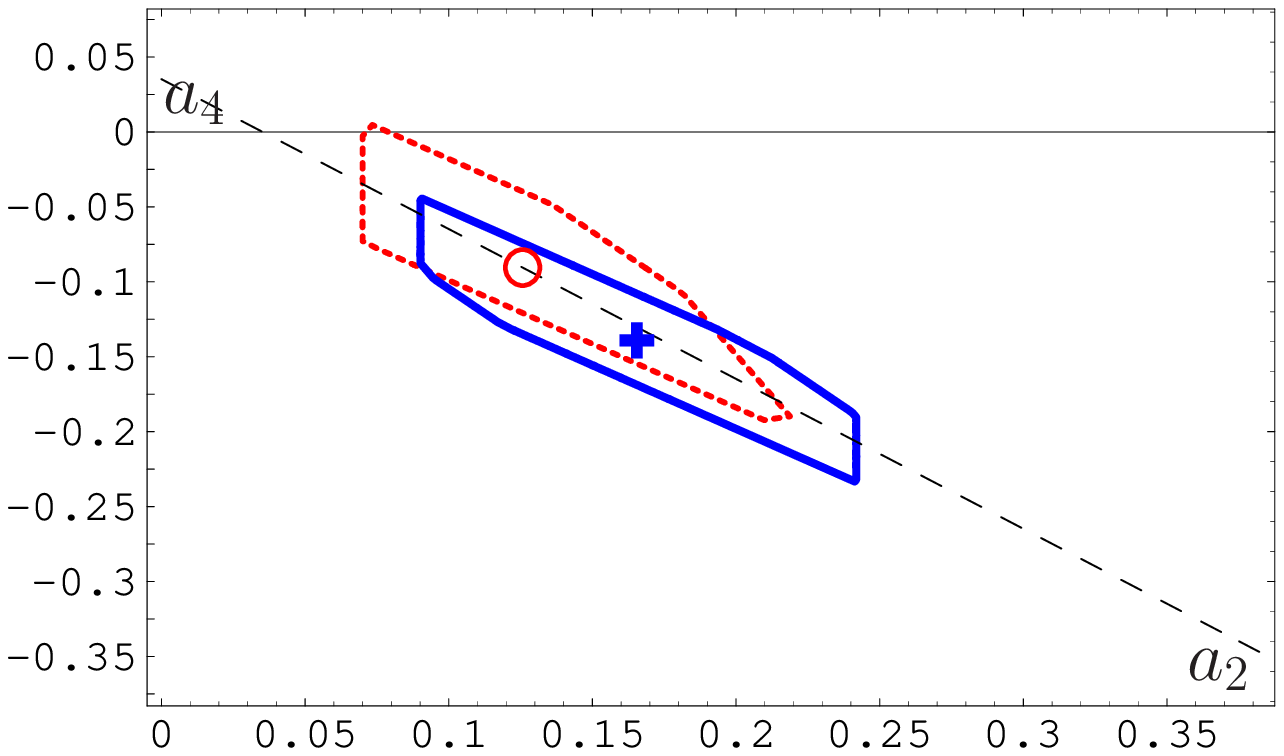}%
        ~~~~~\includegraphics[width=0.47\textwidth]{
              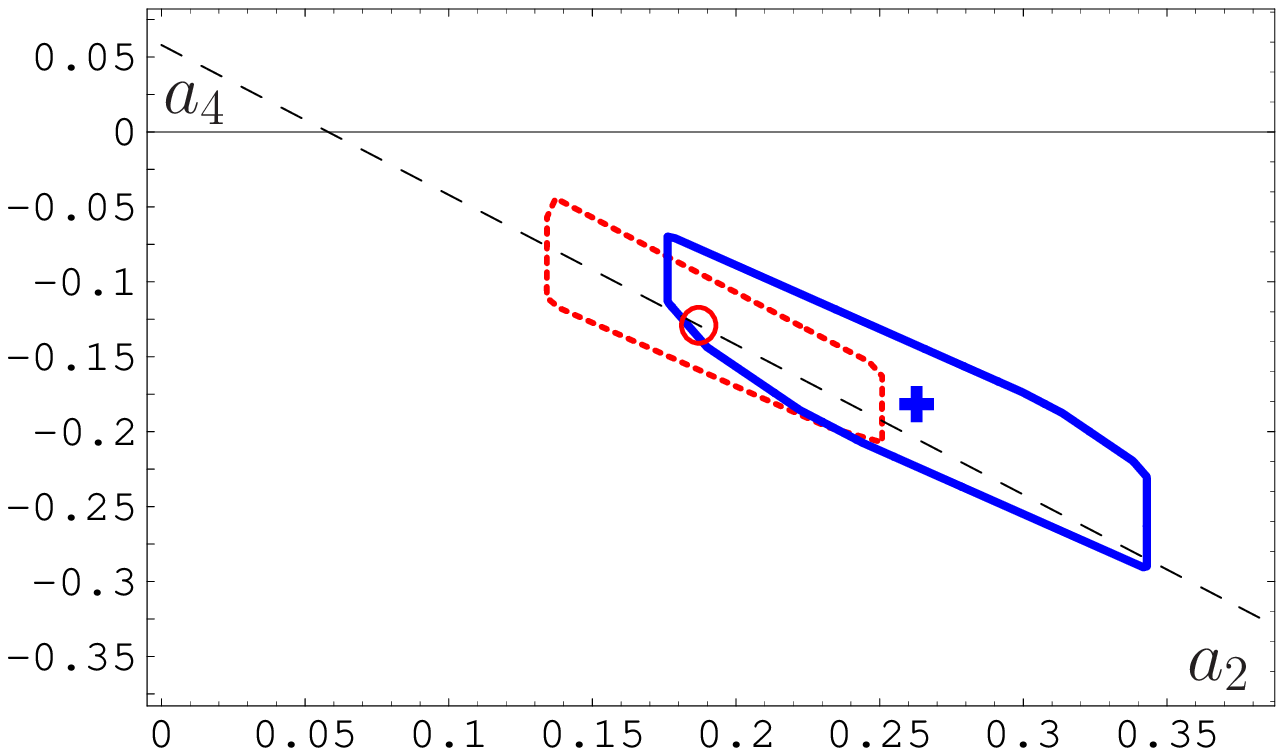}}
  \caption{\label{fig:Pion.DA.Bunch} 
   Allowed values of the pion DA parameters $a_2$ and $a_4$ 
   are bounded by the solid blue line. 
   Region bounded by the dotted red line represents results obtained 
   in the minimal model~\protect\cite{BMS01}.
   Left panel show the results for the value $\lambda_q^2=0.5$ GeV$^2$,
   right panel -- for the value $\lambda_q^2=0.4$~GeV$^2$.
    All values are normalized at $\mu^2=1.35~\text{GeV}^2$.}
\end{figure}

The obtained QCD vacuum model allows us to calculate 
moments of the pion DA $\varphi_{\pi}(x,\mu^2)$~\cite{Rad77} 
more accurately.
\begin{equation}
 \label{eq:Pion.DA.Def}
  \langle{0\mid \bar d(z)\gamma^{\mu}\gamma_5 u(0)\mid \pi(P)}\rangle\Big|_{z^2=0}
   = i f_{\pi}P^{\mu}
    \int^1_0 dx\,e^{ix(zP)}\ \varphi_{\pi}(x,\mu^2)\ .
\end{equation}
The results of the analysis of $\langle\xi^{2N}\rangle_\pi$
in the NLC QCD sum rules are given in the Table \ref{tab:pion.moments}. 
\begin{table}[hb]
\caption{Pion DA moments $\langle \xi^N \rangle_{\pi}(\mu_0^2)$,
 determined at $\mu_0^2 = 1.35~\text{GeV}^2$.
 \label{tab:pion.moments}}
\begin{ruledtabular}
\footnotesize
\begin{tabular}{c|c|c|c|c|c|c}
 Model
   &
 $f_{\pi}\left(\text{GeV}\right)$
   &$\hspace{0.1mm}N=2\hspace{0.1mm}$
     &$\hspace{0.1mm}N=4\hspace{0.1mm}$
       &$\hspace{0.1mm}N=6\hspace{0.1mm}$
         &$\hspace{0.1mm}N=8\hspace{0.1mm}$
           &$\hspace{0.1mm}N=10\hspace{0.1mm}$
             \\ \hline
 ${\strut\vphantom{\vbox to 6mm{}}}$
 Minimal~\cite{BMS01}
   &
 $0.137(8)$
   & $0.266(20)$
     & $0.115(11)$
       & $0.060(7)$
         & $0.036(5)$
           & $0.025(4)$
             \\ \hline
 ${\strut\vphantom{\vbox to 6mm{}}}$
 Ansatz (\ref{eq:Ansatz.Xv1.W})
   & $\hspace{1mm}0.140(13)\hspace{1mm}$
   & $\hspace{1mm}0.290(29)\hspace{1mm}$
     & $\hspace{1mm}0.128(13)\hspace{1mm}$
       & $\hspace{1mm}0.067(7)\hspace{1mm}$
         & $\hspace{1mm}0.040(5)\hspace{1mm}$
           & $\hspace{1mm}0.025(4)\hspace{1mm}$
             \\
\end{tabular}
\end{ruledtabular}
\end{table}
One can see from this table
that the values of the pion DA moments in the new Gaussian model of QCD vacuum
are systematically different from those,
corresponding to the minimal model.
Allowed region for the Gegenbauer coefficients $a_2$ and $a_4$ 
are shown in Fig.\ \ref{fig:Pion.DA.Bunch}.
These coefficients define the pion DA 
in a form of the expansion in Gegenbauer polynomials $C^{3/2}_{2n}(2x-1)$, 
being the eigenfunctions of the 1-loop ER-BL~\cite{ER80,LB79} evolution kernel:
\begin{equation}
 \label{eq:DA.2.Geg}
  \varphi_\pi(x;\mu^2=1.35~\text{GeV}^2) 
   = 6\,x\bar{x}\,\Bigl[1+a_2\,C^{3/2}_2(2x-1)+a_4\,C^{3/2}_4(2x-1)\Bigr]\,.
\end{equation}

In order to test the self-consistency of our procedure 
of DA restoration on the basis of information about
its first five conformal moments,
we use the same technique as in~\cite{BM98,BMS01}.
Namely, we construct the special SR 
for the inverse moment $\langle{x^{-1}}\rangle_{\pi}$
and the result of its processing 
$\langle{x^{-1}}\rangle_{\pi}^\text{SR}$ 
is compared with 
the inverse moment obtained from representation (\ref{eq:DA.2.Geg}):
\begin{eqnarray}
 \label{eq:Inv.Mom.2.Geg}
  \langle{x^{-1}}\rangle_{\pi}^\text{DA} 
   &=& 3\,\left(1+a_2+a_4\right)\,.
\end{eqnarray}
For the value $\lambda^2_q=0.4$~GeV$^2$ we get following results:
$$ \langle{x^{-1}}\rangle_\pi^{\text{DA}} = 3.25\pm0.20\,;\quad
   \langle{x^{-1}}\rangle_\pi^{\text{SR}} = 3.40\pm0.34\,,$$
and for the value $\lambda^2_q=0.5$~GeV$^2$ --- these:
$$ \langle{x^{-1}}\rangle_\pi^{\text{DA}} = 3.08\pm0.15\,;\quad
   \langle{x^{-1}}\rangle_\pi^{\text{SR}} = 3.27\pm0.35\,.$$
The obtained inverse moments in both cases 
are in good mutual agreement.
This confirms the self-consistency of the pion DA recovery procedure.
\newpage
\section{Conclusion}
 \label{sec:Conclusion}
Here we considered the Gaussian model of the nonlocal vacuum quark 
and quark-gluon condensates 
in QCD.
We analyzed the Lorenz structure of the correlator $\Pi_{\mu\nu}(q)$ 
of two vector quark currents
and showed that in the minimal Gaussian model 
of the nonperturbative QCD vacuum~\cite{MR89,BM98,BMS01},
this correlator is non-transversal 
and nonlocal condensates do not satisfy QCD equations of motion.

\begin{figure}[b]
 \centerline{\includegraphics[width=0.47\textwidth]{
             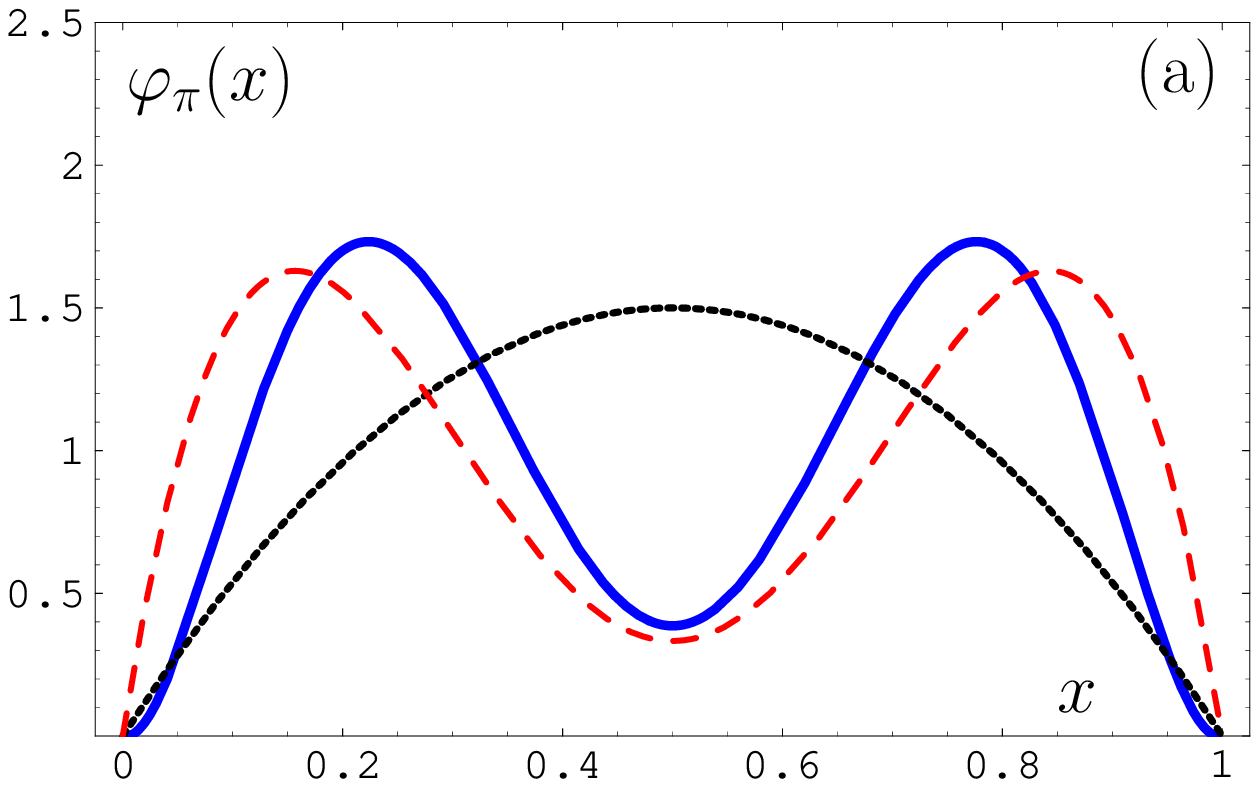}%
          ~~~\includegraphics[width=0.47\textwidth]{
             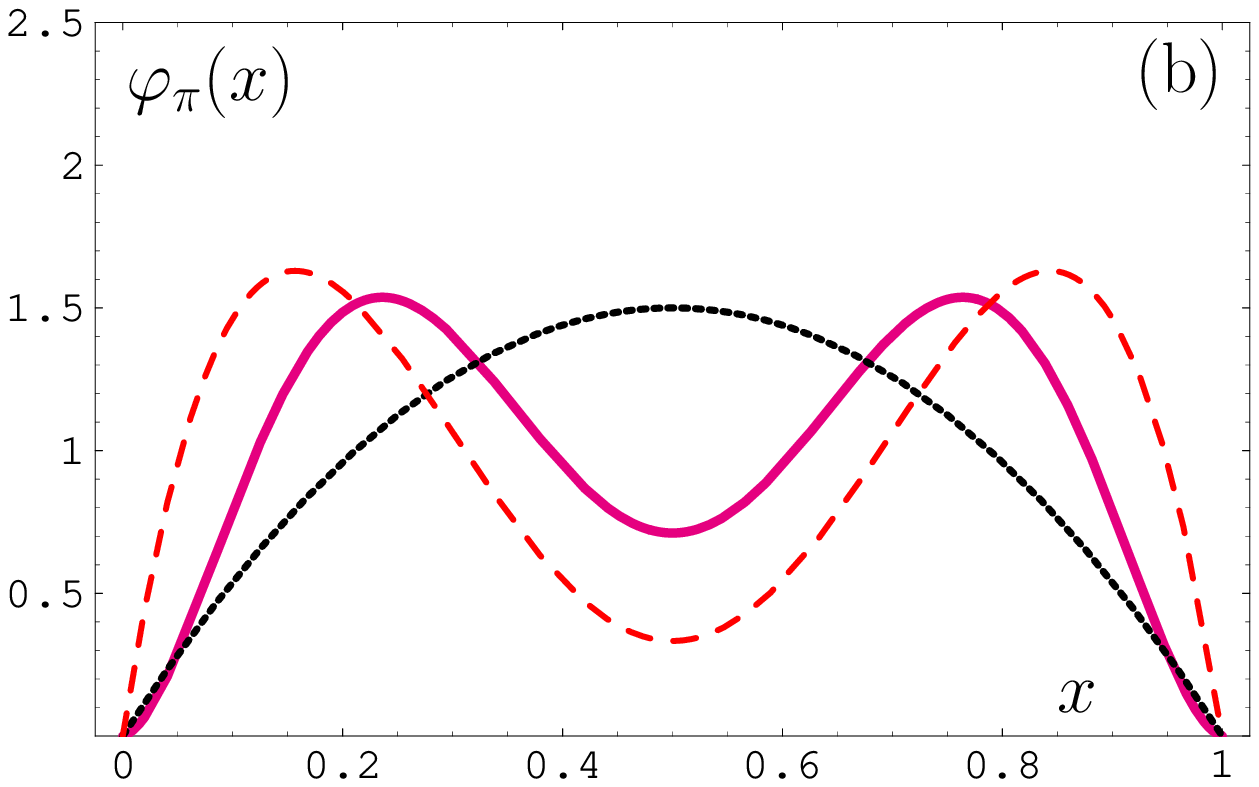}}
  \caption{\label{fig:Pion.Bunch} 
    Profiles of the pion DAs corresponding to the central points of the ``bunches'' 
    for the value of the nonlocality parameter $\lambda_q^2=0.4$~GeV$^2$.
    \textbf{Panel (a)}: The blue solid line represents the result
    obtained in the improved Gaussian model 
    (symbol \BluTn{\ding{57}} on the right part of Fig.\ \protect\ref{fig:Pion.DA.Bunch}).
    \textbf{Panel (b)}: The red solid line represents the result 
    obtained in the minimal Gaussian model (BMS model~\protect\cite{BMS01},
     symbol \RedTn{$\bm{\circ}$} on the right part of Fig.\ \protect\ref{fig:Pion.DA.Bunch}).
    For comparison we show here also the asymptotic DA (dotted line) 
    and Chernyak--Zhitnitsky (CZ) DA~\protect\cite{CZ82} (red dashed line).}
\end{figure}
To ameliorate the situation we suggested  
the improved Gaussian model for nonlocal vacuum quark and quark-gluon condensates
in QCD, 
Eqs.\ (\ref{eq:Delta.Ansatz.SV}) and (\ref{eq:3L.D.A.Deriv}).
This model satisfies QCD equations of motion for quark fields
and the revealed breakdown of gauge invariance 
is minimized by the special choice of parameters,
see Eqs.\ (\ref{eq:Ansatz.Xv1.W}).

Using this improved model of the nonlocal QCD vacuum
we analyzed QCD SRs for the pion DA.  
We revealed that in the new QCD vacuum model 
the NLC SRs produce again a  2-parameter 
``\textit{bunch}'' of admissible DAs.
The allowed values of this bunch parameters $a_2$ and $a_4$
are shown in Fig.\ \ref{fig:Pion.DA.Bunch}. 
These models are in a good agreement with the results 
of independent SR for the pion DA inverse moment,
$\langle{x^{-1}}\rangle_\pi^{\text{SR}}$.

We emphasize here 
that obtained earlier 
in the minimal Gaussian model of QCD vacuum 
the BMS model~\cite{BMS01},
shown in Fig.\ \ref{fig:Pion.DA.Bunch} by symbol \RedTn{$\bm{\circ}$},
is inside the allowed region dictated by the improved QCD vacuum model.
This testifies to the heredity of both Gaussian models,
the minimal one and the improved one.
Moreover, that also means 
that all the characteristic features of the BMS bunch 
are valid also for the improved bunch:
one can see in Fig.\ \ref{fig:Pion.Bunch}
that in comparison with the CZ model~\cite{CZ82} 
(dashed red line, $a_2=0.52$ and $a_4=0$ at $\mu^2=1.35$~GeV$^2$)
the NLC-dictated models are much more end-point suppressed,
although are double-humped.

This results in completely different values of the inverse moment:
$\langle{x^{-1}}\rangle_\pi^{\text{CZ}}=4.56$,
whereas in our case 
$\langle{x^{-1}}\rangle_\pi = 3.24\pm0.20$.\footnote{%
Note that smaller errors in the analysis~\cite{BMS01} are related
with their estimation only by stability with respect to Borel parameter $M^2$ 
variation inside the ``fidelity window'' of QCD SRs.
In our paper we also take into account internal errors of the QCD SR approach
and suggest that the overall error can not be smaller than 10\%.}

\acknowledgments
We would like to thank A.~Dorokhov, S.~Mikhailov, N.~Stefanis and O.~Teryaev
for stimulating discussions and useful remarks.
This work was supported in part by the Heisenberg--Landau Programme,
grant 2006, and the Russian Foundation for Fundamental Research, 
grant No.\ 06-02-16215.

\newpage
\begin{appendix}
\appendix
 \textbf{\large Appendix}
  \vspace{-3mm}
\setcounter{section}{0}
\section{$O(\alpha_s\langle{\bar{\psi}\psi}\rangle^2)$-contributions}
 \renewcommand{\theequation}{\thesection.\arabic{equation}}
  \label{App:A.1}\setcounter{equation}{0}
$O(\alpha_s\langle{\bar{\psi}\psi}\rangle^2)$-order terms of
 $\Delta_\text{2V}\Pi^N_{\mu\nu}$ are determined by four objects.
They are bilocal vector condensate($\Delta_\text{2V}\Pi^N_{\mu\nu}$),
3-local quark-gluon-antiquark condensate ($\Delta_{\bar{q}Aq}\Pi^N_{\mu\nu}$),
4-quark condensates ($\Delta_\text{4Q$_1$}\Pi^N_{\mu\nu}$ and $\Delta_\text{4Q$_2$}\Pi^N_{\mu\nu}$).
We consider terms for diagrams Figs.\,\ref{fig:NLC1} and \ref{fig:NLC2}.
Contribution of mirror-conjugate diagrams are taken into account by symmetrical consideration,
see Appendix \ref{App:A.2}.
\begin{eqnarray}
 \Delta_\text{2V}\Pi^N_{\mu\nu}
        &=& \frac{i}{(nq)^N}\int\!\!dx\,e^{iqx}
               \langle{\bar{u}(0)\gamma_\mu
                    (-in\nabla_0)^Nd(0)\BraSquare{-3}{14}{90}{14}\bar{d}(x)
                       \gamma_\nu u(x)}\rangle;~~~\label{eq:Delta.PiN.2V}\\
 \Delta_{\bar{q}Aq}\Pi^N_{\mu\nu}
        &=& \frac{i(ig)}{(nq)^{N}}
             \int\!\!dx\,e^{iqx}\!\!
              \int\!\!dy
              \langle{\bar{d}(0)\gamma_\mu
                    (-in\nabla_0)^Nu(0)\BraSquare{-3}{14}{90}{14}\bar{u}(y)
                     \gamma_\rho\widehat{A}_\rho(y)
                     u(y)\BraSquare{-3}{14}{90}{14}\bar{u}(x) 
                       \gamma_\nu d(x)}\rangle;
                        ~~~~~\label{eq:Delta.PiN.qAq}\\
  \Delta_\text{4Q$_1$}\Pi^N_{\mu\nu}
        &=& \frac{i(ig)^2}{(nq)^{N}}
             \int\!\!dx\,e^{iqx}\!\!
              \int\!\!dy\!\!
               \int\!\!dz\nonumber\\
        &&\times\langle{%
     \bar{d}(0)\gamma_\mu(-in\overrightarrow{\nabla}_0)^Nu(0)
      \BraSquare{-3}{14}{90}{14}
       \bar{u}(y)\gamma^\rho\widehat{A}_\rho(y)
        \BraSquare{50}{18}{90}{74}
         u(y)\bar{u}(x)\gamma_\nu d(x)
          \BraSquare{-3}{14}{90}{14}
           \bar{d}(z)\gamma^\lambda\widehat{A}_\lambda(z)d(z)}\rangle;
            ~~~~~\label{eq:Delta.PiN.4q1}\\
 \Delta_\text{4Q$_2$}\Pi^N_{\mu\nu}
        &=& \frac{i(ig)^2}{(nq)^{N}}
             \int\!\!dx\,e^{iqx}\!\!
              \int\!\!dy\!\!
               \int\!\!dz\nonumber\\
        &&\times\langle{
           \bar{d}(0)
                    (-in\overrightarrow{\nabla}_0)^N\gamma_\mu u(0)
                     \BraSquare{-3}{14}{90}{14}
          \bar{u}(y)\gamma^\rho\widehat{A}_\rho(y)
                  \BraSquare{22}{18}{90}{48}
          u(y)\bar{u}(z)\gamma^\lambda \widehat{A}_\lambda(z)u(z)
                      \BraSquare{-3}{14}{90}{14}
          \bar{u}(x)\gamma_\nu u(x)}\rangle.
           ~~~~~\label{eq:Delta.PiN.4q2}
\end{eqnarray}
Values $\Delta_{k}\Pi_\text{L}^N(M^2)$, $k=2$V, $\bar{q}Aq$, 4Q$_1$ и 4Q$_2$ 
are defined in (\ref{eq:2.V.L})--(\ref{eq:4.q.2.L}), where
\begin{eqnarray}
H_1    &=&
           N(N+1)\bar{\Delta}_1\Delta_2^2
           +H_2\frac{\Delta_2}{\bar{\Delta}_1}-NH_3
           \,,\nonumber\\
H_2    &=&
           -\bar{\Delta}_1\left((N+3)\Delta_1\Delta_2\left(\bar{\Delta}_1-\Delta_3\right)
           +\Delta_3\left(3\Delta_2+2\Delta_1\bar{\Delta}_1\right)\right)
               \,,\nonumber\\
H_3    &=&
          \Delta_2\left(\left(N+(N+3)\Delta_1\right)\Delta_2\bar{\Delta}_1
          +\Delta_3\left(3\Delta_2+\Delta_1 \bar{\Delta}_1-(N+3)\Delta_1\Delta_2\right)\right)
              \,,\nonumber\\
G_1    &=& 
          -N(N+1)\Delta_2^2\bar{\Delta}_1\left(\bar{\Delta}_1-\Delta_2\right)^2
          +G_2\frac{\Delta_2}{\bar{\Delta}_1}-G_3N
          \,,\nonumber\\
G_2    &=&
          \bar{\Delta}_1^2\Delta_2\left[3(N+1)(N+2)\Delta_2^2
          -(N+1)(N+3)\Delta_1\Delta_2^2\right.~~~\nonumber\\&&\left.
          ~~~~~~~~~~~~~~~~~~~~~~~~~~~~~~~~~+N(N+3)\Delta_1\bar{\Delta}_1\Delta_2
          +(N+3)\Delta_1\bar{\Delta}_1^2\right]~~~\nonumber\\&&
          +\bar{\Delta}_1\Delta_3\left(\bar{\Delta}_1-\Delta_2\right)\left[(N+1)\left(\bar{\Delta}_1+2\right)\Delta_2^2
          \right.~~~\nonumber\\&&\left.
          ~~~~~~~~~~~~~~~~~~~~~~~~~~~~~~~~~
          +(N-1)\Delta_1\bar{\Delta}_1\Delta_2+3\bar{\Delta}_1\Delta_2+2\Delta_1\bar{\Delta}_1^2\right]
               \,,\nonumber\\
G_3    &=&
          -\Delta_2\left(\bar{\Delta}_1-\Delta_2\right)\left[\Delta_1\Delta_3\left(\bar{\Delta}_1-\Delta_2\right)^2
          +3\Delta_2\Delta_3\left(\bar{\Delta}_1-\Delta_2\right)\right.\nonumber\\&&\left.
          +\Delta_2\bar{\Delta}_1\left(N\bar{\Delta}_1
          +(N+3)\left(\Delta_1\bar{\Delta}_1+\Delta_2\left(\bar{\Delta}_1+1\right)\right)\right)\right]
              \,,\nonumber \\
F_1    &=&
          \left(n+1+\bar{\Delta}\right)(n+2) (n+3) 
           \,,~~~
F_2    = 
            \bar{\Delta }-(n+3) [(n+1) (n+4) \Delta +1]
           \,,\nonumber\\
F_3    &=&
           (n+3) \bar{\Delta }-1
           \,,~~~\nonumber
\end{eqnarray}
and $\Delta=\Lambda_S/M^2$, $\bar{\Delta}=1-\Delta$,  
  $\Delta_i=\alpha_i/M^2$, $\bar{\Delta}_1=1-\Delta_1$.

For transverse components $\Delta_{k}\Pi_\text{T}^N(M^2)$,
 see \ (\ref{eq:Delta.Pi.T.k}), 
  corresponding value are given by following expressions. 
\begin{eqnarray*}
 \widetilde{\varphi}(\alpha_1,\alpha_2,M^2)
 &=& \frac{x\,\theta \left(\Delta_1-\bar{x}\right)}
          {\Delta _1^2 \Delta _2 \bar{\Delta }_1^2}
      \left(\bar{x}\Delta_2\bar{\Delta}_1
           +\log\left(\frac{x\Delta_1\bar{\Delta}_2}
                           {x\Delta _1-(\Delta_1-\bar{x})\Delta_2}
                \right)
                 \Delta_1(\Delta_1-\bar{x})\bar{\Delta}_2
      \right);\\
 \widetilde{\varphi}_{1}(\alpha_1,\alpha_2,\alpha_3,M^2)
 &=& \left(\frac{\Delta_3}{\Delta_2}
          -\frac{\bar{\Delta}_1}{\Delta _2}
     \right)\delta\left(\bar{x}-\Delta_1\right)
   - \left(1
          -\frac{\bar{\Delta}_1}{\Delta _2}
     \right)\delta\left(\bar{x}-\Delta _1-\Delta_2\right)~\\ 
 &-&
     \frac{x\left(x\Delta_3
                 +\Delta _2\left(\Delta_1+\Delta_3-1\right)
            \right)}
           {\bar{\Delta}_1^2\Delta _2^2}\,
       \theta\left(\bar{x}-\Delta_1\right)
        \theta\left(\Delta_1+\Delta_2-\bar{x}\right)\,;~\\
\widetilde{\varphi}_{2}(\alpha_1,\alpha_2,\alpha_3,M^2)
 &=& -\left(1
           -\frac{\bar{\Delta}_1}{\Delta_2}
      \right)\delta\left(\bar{x}-\Delta_1-\Delta_2\right)~\\ 
 &+& \frac{x\left(2\left(\Delta_1-\bar{x}\right)\Delta_3
                 +\Delta_2\left(\Delta_1+\Delta_3-1\right)
            \right)}
          {\bar{\Delta}_1\Delta_2^3}\,
       \theta\left(\bar{x}-\Delta_1\right)
        \theta\left(\Delta_1+\Delta_2-\bar{x}\right)\,;~\\
\widetilde{\varphi}_{3}(\alpha_1,\alpha_2,\alpha_3,M^2)
 &=& -\frac{x\left(\left(\Delta_1-\bar{x}\right)\Delta_3
                  +\Delta_2\left(\Delta_1+\Delta_3-1\right)
             \right)}
           {\bar{\Delta}_1^2\Delta_2^2}\,
       \theta\left(\bar{x}-\Delta_1\right)
        \theta\left(\Delta_1+\Delta_2-\bar{x}\right)\,,~
\end{eqnarray*}
where $\Delta_i=\alpha_i/M^2$, $\bar{\Delta}_i=1-\Delta_i$
and $\bar{x}=1-x$.

Отметим, что результат представлен для параметрических функций
$f_i(\alpha_1,\alpha_2,\alpha_3)$ и $f_S(\alpha)$ таких, 
что при интегрировании 
вклад дают области 
$\alpha_1+\alpha_2<M^2$, 
$\alpha_1+\alpha_3<M^2$ и 
$2\,\alpha<M^2$.
Для анзаца (\ref{eq:Delta.Ansatz.SV}), (\ref{eq:3L.D.A.Deriv}) 
эти условия соответствуют области, 
в которой работают ПС КХД.
Note that result are presented
 for parametric functions $f_i(\alpha_1,\alpha_2,\alpha_3)$ and $f_S(\alpha)$
  such that only $\alpha_1+\alpha_2<M^2$, 
$\alpha_1+\alpha_3<M^2$ 
$2\,\alpha<M^2$ integration domains give contribution.
For Ansatz (\ref{eq:Delta.Ansatz.SV}), (\ref{eq:3L.D.A.Deriv})
this conditions correspond to working area of QCD sum rules.
\section{Conformal moments}
 \renewcommand{\theequation}{\thesection.\arabic{equation}}
  \label{App:A.2}\setcounter{equation}{0}
Let us consider linear combinations of moments $\Delta\Pi_\text{L}^N$, 
\begin{eqnarray}
\nonumber
  \Delta\langle\xi^{2N}\rangle_\text{L}
    \equiv
      \int\limits_0^1(2x-1)^{2N}\varphi(x)\,dx
    = \sum_{k=0}^{2N}(-2)^{2N-k}{2N\choose k}
       \int\limits_0^1\!x^{2N-k}\varphi(x)\,dx\,.
\end{eqnarray}
This combinations is named conformal moments. 
Namely this moments are analyzed in QCD sum rules for meson DA.

Reflection-symmetrical diagrams are equal to calculated diagrams.
If $x$-density for calculated diagrams are $\varphi_0(x)$,
  then $\varphi_0(1-x)$ are density for M.~C. diagrams and 
\begin{eqnarray}\nonumber%
  \int\limits_0^1(2x-1)^{2N}
   \varphi_0(x)\,dx
  = \int\limits_0^1(2x-1)^{2N}
     \varphi_0(1-x)\,dx\,.
\end{eqnarray}
That is full contribution in conformal moment of fixed diagram
 is equal to doubled term of either of diagram.
\begin{eqnarray}
 \nonumber
  \Delta\langle\xi^{2N}\rangle_\text{L}
   = 2\,\int\limits_0^1(2x-1)^{2N}
      \varphi_0(x)\,dx\,.    
\end{eqnarray}
Denoting
\begin{eqnarray}
\nonumber
  \Delta\widetilde{\Pi}_0^k
   \equiv \int\limits_0^1\!x^k
           \varphi_0(x)\,dx\,,
\end{eqnarray}
we immediately obtain the necessary conformal moments
\begin{eqnarray}
\nonumber
 \Delta\langle\xi^{2N}\rangle_\text{L}
  = 2\,\sum_{k=0}^{2N}(-2)^{k}{2N\choose k}
     \Delta\widetilde{\Pi}_0^k\,.
\end{eqnarray}

\end{appendix}


\end{document}